\documentclass[12pt]{article}
\usepackage{amsmath,amsthm}
\usepackage{rotating}

\def\pbnr{}
\def\speaker{Chengdong Fu}
\def\onbehalfof{BESIII collaboration}
\def\title{Dalitz Plot Analysis of the $D^+\rightarrow K^0_S \pi^+ \pi^0$ Decay}
\def\affiliation{Institute of High Energy Physics, Beijing, China}
\def\support{E-mail: fucd@ihep.ac.cn}

\newcommand\pubnumber{\pbnr}
\newcommand\pubdate{\today}
\def\Title#1{\begin{center} {\Large #1 } \end{center}}
\def\Author#1{\begin{center}{ \sc #1} \end{center}}

\newcommand{\OnBehalf}[1]{\sbox0{#1}\ifdim\wd0=0pt
        {}
	\else
	{\\on behalf of #1}
	\fi}
\newcommand{\SupportedBy}[1]{\sbox0{#1}\ifdim\wd0=0pt
        {}
	\else
	{\footnote{#1}}
	\fi}
\def\Address#1{\begin{center}{ \it #1} \end{center}}

\newcommand\pubblock{\includegraphics[width=5cm]{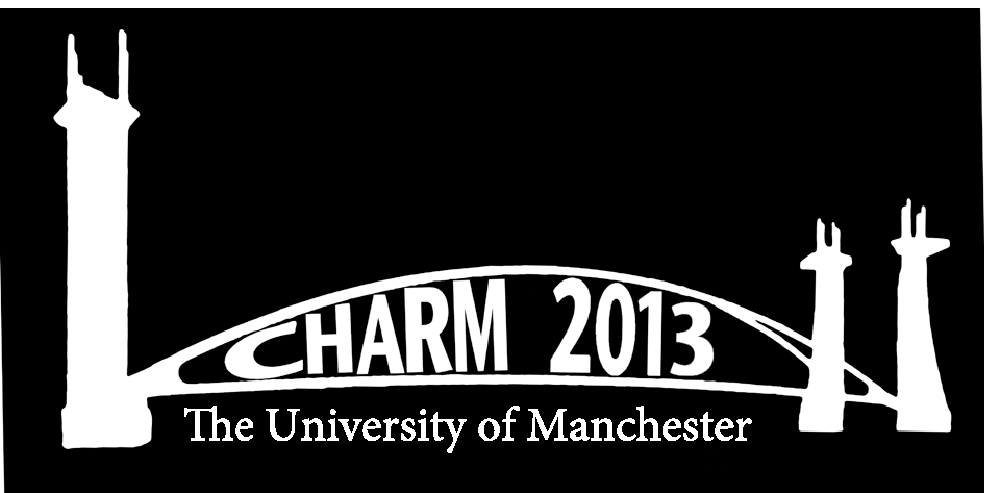}\hfill{\begin{tabular}{l} \pubnumber\\
         \pubdate  \end{tabular}}}
\newenvironment{Abstract}{\begin{quotation}  }{\end{quotation}}
\newenvironment{Presented}{\begin{quotation} \begin{center} 
             PRESENTED AT\end{center}\bigskip 
      \begin{center}\begin{large}}{\end{large}\end{center} \end{quotation}}
\def\Acknowledgements{\bigskip  \bigskip \begin{center} \begin{large}
             \bf ACKNOWLEDGEMENTS \end{large}\end{center}}
\def\venue{The 6$^{th}$ International Workshop on Charm Physics\\
(CHARM 2013)\\
Manchester, UK,  31 August -- 4 September, 2013}

\def\beq{\begin{equation}}
\def\eeq#1{\label{#1}\end{equation}}
\def\eeqn{\end{equation}}

\def\beqa{\begin{eqnarray}}
\def\eeqa#1{\label{#1}\end{eqnarray}}
\def\eeqan{\end{eqnarray}}

\let\bar=\overbar

\def\Dslash{\not{\hbox{\kern-4pt $D$}}}
\def\dslash{\not{\hbox{\kern-2pt $\del$}}}

\def\msb{{\bar{\ssstyle M \kern -1pt S}}}

\begin{document}
\begin{titlepage}
\pubblock

\vfill
\Title{\title}
\vfill
\Author{\speaker\SupportedBy{\support}\OnBehalf{\onbehalfof}}
\Address{\affiliation}
\vfill
\begin{Abstract}
We perform an analysis of the $D^+\rightarrow K^0_S \pi^+ \pi^0$ Dalitz plot using a data set of 2.92~fb$^{-1}$ of $e^+e^-$ collisions at the $\psi(3770)$ accumulated by the BESIII Experiment, in which 166694 candidate events are selected with a background of 15.1\%.  The Dalitz plot is found to be well-represented by a combination of six quasi-two-body decay channels ($K^0_S\rho^+$, $K^0_S\rho(1450)^+$, $\overline{K}^{*0}\pi^+$, $\overline{K}_0(1430)^0\pi^+$, $\overline{K}(1680)^0\pi^+$, $\overline{\kappa}^0\pi^+$) plus a small non-resonant component. We also consider a model-independent approach to confirm an obvious phase shift for the $\overline{\kappa}\pi$ component. Using the fit fractions from this analysis,  partial branching ratios are updated with higher precision than previous measurements.
\end{Abstract}
\vfill
\begin{Presented}
\venue
\end{Presented}
\vfill
\end{titlepage}
\def\thefootnote{\fnsymbol{footnote}}
\setcounter{footnote}{0}
%

\section{Introduction}

A clear understanding of final-state interactions in exclusive weak decays is an important ingredient in our ability to predict decay rates and to model the dynamics of two-body decays of charmed mesons. Final-state interactions can cause significant changes in decay rates, and can cause shifts in the phases of decay amplitudes. Clear experimental measurements can help refine theoretical models of these phenomena.

Three-body decays provide a rich laboratory in which to study the interferences between intermediate-state resonances.  They also provide a direct probe of final-state interactions in certain decays. When a particle decays into three pseudo-scalar particles, intermediate resonances dominate the decay rate and amplitudes are typically obtained with a Dalitz plot analysis technique \cite{Dalitz}. This provides the opportunity to experimentally measure both the amplitudes and phases of the intermediate decay channels, which in turn allows us to deduce their relative branching fractions. These phase differences can even allow details about very broad resonances to be extracted by observing their interference with other intermediate states.

A large contribution from a $K\pi$ $S$-wave intermediate state has been observed in earlier experiments. Both E791 \cite{E791b} and CLEO-c \cite{CLEOc} interpreted their data with a Model-Independent Partial Wave Analysis (MIPWA) and found a phase shift at low $K\pi$ mass to confirm the $\kappa\pi$ component in the $D^+ \rightarrow K^- \pi^+ \pi^+$ decay. Complementary to this channel, the $D^+\rightarrow K^0_S \pi^+\pi^0$ decay is also a golden channel to study the $K\pi$ $S$-wave in $D$ decays.

The previous Dalitz plot analysis of  $D^+\rightarrow K^0_S \pi^+\pi^0$ by MARKIII \cite{MARK3} included only two intermediate decay channels,  $K^0_S\rho$ and $\overline{K}^{*0}\pi^+$, and was based on a small data set. A much larger data sample of e$^+$e$^-$ collisions at $\sqrt{s} \approx 3.773$ GeV has been accumulated with the BESIII detector \cite{BES3} running at the Beijing Electron-Positron Collider (BEPCII) \cite{BEPC2}. With much larger statistics, it is possible to measure relative branching fractions more precisely and to find more intermediate resonances.

\section{Dalitz Fit at BESIII}

BESIII has established the Dalitz plot analysis based on the technology of maximum likelihood fit. The likelihood function is defined as $\mathcal{L}=\prod^{N}_{i=1} \mathcal{P}(x_i,y_i)$, where $N$ is the event number and $\mathcal{P}(x,y)$ is the probability density function on Dalitz plot. For signal with background in data, it is described as
\begin{equation}
\mathcal{P}(x,y)= f_{S}\frac{|\mathcal{M}(x,y)|^2\varepsilon(x,y)}{\int\limits_{DP}|\mathcal{M}(x,y)|^2\varepsilon(x,y)dxdy} +f_{B}\frac{B(x,y)}{\int\limits_{DP}B(x,y)dxdy},
\label{eqn:probability}
\end{equation}
where $\mathcal{M}(x,y)$ is the decay matrix element, $\varepsilon(x,y)$ is the efficiency shape, $B(x,y)$ is the background shape, $f_{S}$ and $f_{B}$ are the fractions of signal and background, respectively. The $DP$ denotes the kinematic limit on the Dalitz ploy. The decay matrix element is contributed by isobar model. The efficiency is parameterized by Monte-Carlo sample \cite{BES3mc}. The background includes two parts: peaking background and non-peaking background. The peaking background is estimated by Monte-Carlo simulation, and the non-peaking background is parameterized by the low and high sidebands of the distribution of the recoiling mass of selected $D$ meson $m_{rec}$ of data. The fractions of signal and background are fitted by the distribution of the $m_{rec}$.

Because the high-mass $m_{rec}$ sideband has a significant contribution from signal events due to a tail caused by initial state radiation, we consider a contribution of signal for these events. The contribution of signal is obtained by fitting on data using the low sideband only as background approximation. The background process and the signal process are repeated to approximate the expected more and more.

\section{Results of the D to Ks pi pi0 Decay}
\newcommand{\grad}{\ensuremath{^{\circ}}}
\begin{table}[ht]
  \begin{center}
  \begin{tabular}{|l|c|c|c|c|c|}
    \hline
    \hline
    Decay Mode & Par. & Flavor & w/o $\overline{\kappa}$ & w/o NR & Final \\
    \hline
    \hline
    Non-resonant                    & FF(\%) & 4.5$\pm$0.7     & 18.3$\pm$0.6    &                 & 4.6$\pm$0.7   \\
                                    & ~$\phi$(\grad) & 269$\pm$6       & 232.7$\pm$1.3   &                 & 279$\pm$6     \\
    $K^0_S\rho(770)^+$              & FF(\%) & 84.6$\pm$1.8    & 82.0$\pm$1.3    & 86.7$\pm$1.1    & 83.4$\pm$2.2  \\
                                    & ~$\phi$(\grad) & 0(fixed)        & 0(fixed)        & 0(fixed)        & 0(fixed)      \\
    $K^0_S\rho(1450)^+$             & FF(\%) & 1.80$\pm$0.20   & 6.03$\pm$0.29   & 0.63$\pm$0.12   & 2.13$\pm$0.22 \\
                                    & ~$\phi$(\grad) & 198$\pm$4       & 167.1$\pm$2.1   & 186$\pm$8       & 187.0$\pm$2.6 \\
    $\overline{K}^*(892)^0\pi^+$    & FF(\%) & 3.22$\pm$0.14   & 2.99$\pm$0.10   & 3.30$\pm$0.10   & 3.58$\pm$0.17 \\
                                    & ~$\phi$(\grad) & 294.7$\pm$1.3   & 279.3$\pm$1.2   & 292.3$\pm$1.5   & 293.2$\pm$1.3 \\
    $\overline{K}^*(1410)^0\pi^+$   & FF(\%) & 0.12$\pm$0.05   & 0.18$\pm$0.05   & 0.12$\pm$0.05   &               \\
                                    & ~$\phi$(\grad) & 228$\pm$9       & 301$\pm$10      & 243$\pm$12      &               \\
    $\overline{K}^*_0(1430)^0\pi^+$ & FF(\%) & 4.5$\pm$0.6     & 10.5$\pm$1.3    & 3.6$\pm$0.5     & 3.7$\pm$0.6   \\
                                    & ~$\phi$(\grad) & 319$\pm$5       & 306.2$\pm$2.0   & 317$\pm$4       & 334$\pm$5     \\
    $\overline{K}^*_2(1430)^0\pi^+$ & FF(\%) & 0.118$\pm$0.018 & 0.086$\pm$0.014 & 0.111$\pm$0.015 &               \\
                                    & ~$\phi$(\grad) & 273$\pm$7       & 265$\pm$9       & 267$\pm$7       &               \\
    $\overline{K}^*(1680)^0\pi^+$   & FF(\%) & 0.21$\pm$0.06   & 0.58$\pm$0.08   & 0.43$\pm$0.10   & 1.27$\pm$0.11 \\
                                    & ~$\phi$(\grad) & 243$\pm$6       & 284$\pm$4       & 234$\pm$5       & 251.8$\pm$1.9 \\
    $\overline{K}^*_3(1780)^0\pi^+$ & FF(\%) & 0.034$\pm$0.008 & 0.055$\pm$0.008 & 0.037$\pm$0.008 &               \\
                                    & ~$\phi$(\grad) & 130$\pm$12      & 113$\pm$9       & 131$\pm$11      &               \\
    $\overline{\kappa}^0\pi^+$                 & FF(\%) & 6.8$\pm$0.7     &                 & 18.8$\pm$0.5    & 7.7$\pm$1.2   \\
                                    & ~$\phi$(\grad) & 92$\pm$6        &                 & 11.6$\pm$1.9    & 93$\pm$7      \\
    \hline
    NR+$\overline{\kappa}^0\pi^+$   & FF(\%) & 18.1$\pm$1.4    & 18.3$\pm$0.6    & 18.8$\pm$0.5    & 19.2$\pm$1.8  \\
    $K^0_S\pi^0$ $S$ wave           & FF(\%) & 18.9$\pm$1.0    & 15.8$\pm$1.0    & 21.2$\pm$1.0    & 17.1$\pm$1.4  \\
    \hline
    \hline
  \end{tabular}
  \caption{The preliminary results of the fits to the $D^+ \rightarrow K^0_S\pi^+\pi^0$ Dalitz plot with statistical errors only for different resonance choices, fit fraction (FF) and phase ($\phi$). The \textquotedblleft Final\textquotedblright are momoentum-dependent corrected.}
  \label{tab:fitresult}
  \end{center}
\end{table}
Based on 166694 selected candidate events with a background of 15.1\%, a decay matrix element is constructed by possible intermediate resonance decay modes. After more possible intermediate resonance decay modes were considered in different isobar models, three models are compared principally, the Cabbibo favored model, the model without the $\overline{\kappa}$ and the model without the non-resonant. The results are listed in the column \textquotedblleft Favored\textquotedblright, \textquotedblleft w/o $\overline{\kappa}$\textquotedblright and \textquotedblleft w/o NR\textquotedblright of Table \ref{tab:fitresult}, respectively. It is found that the goodness of fit in the \textquotedblleft w/o $\overline{\kappa}$\textquotedblright model is much worse than in the favored model, which indicates the $\overline{\kappa}$ has a large confidence level in our data. If non-resonant removed, the goodness of fit also becomes worse, indicating that a non-resonant component is indeed present in our data.

\begin{figure}[htb]
  \centering
  \includegraphics[width=0.8\linewidth]{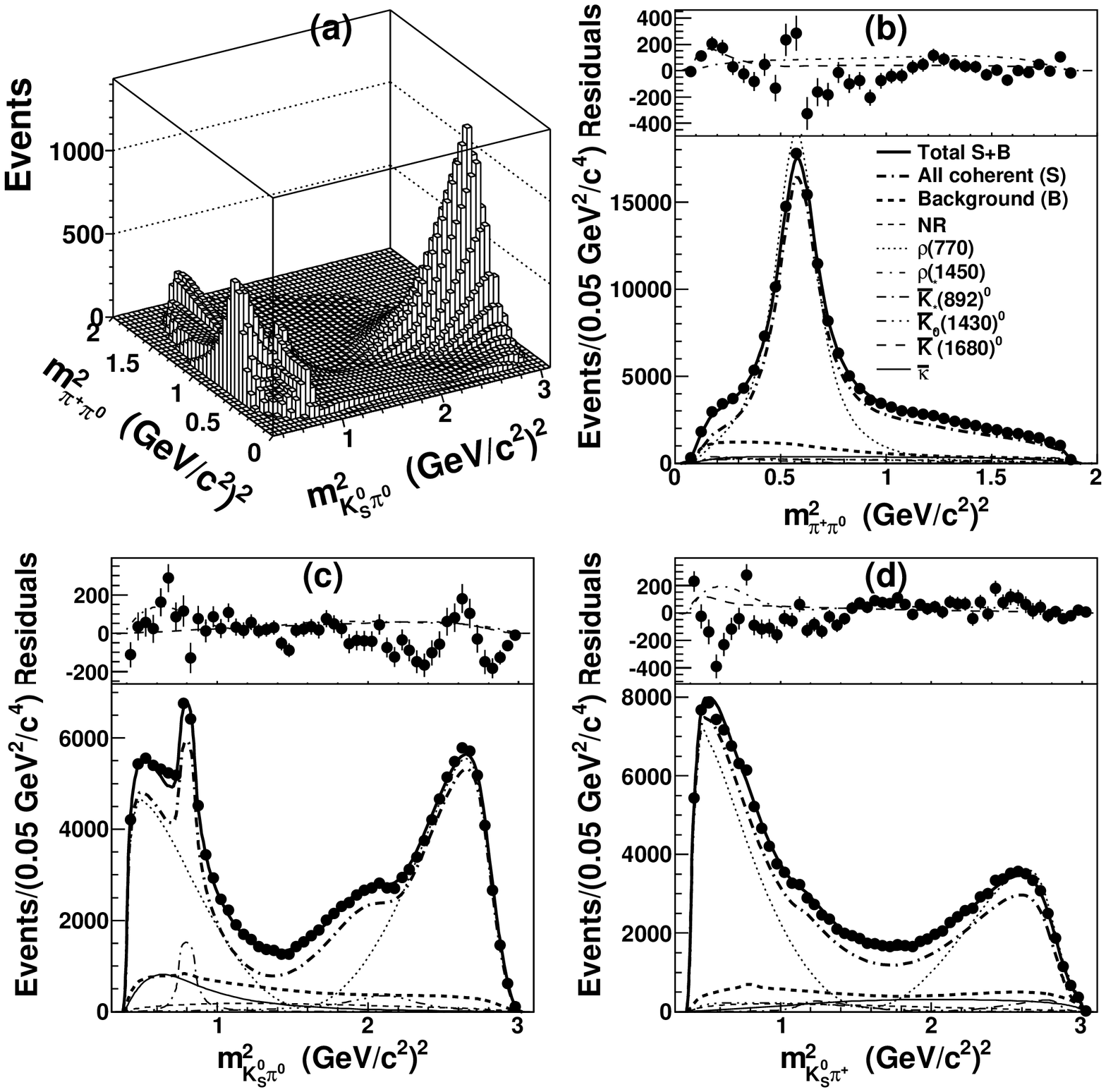}
  \begin{rotate}{45}
  \put(-100,250){\huge{BESIII Preliminary}}
  \end{rotate}
  \caption{The results of fitting the $D^+\rightarrow K^0_S \pi^+\pi^0$ data with final chosen resonances. (a) Distribution of fitted p.d.f. and projections on (b) $m^2_{\pi^+\pi^0}$, (c) $m^2_{K^0_S\pi^0}$ and (d) $m^2_{K^0_S\pi^+}$. Residuals between data and the total p.d.f. are shown by dots with statistical error bars on the top insets.}
  \label{fig:final}
\end{figure}

In the above three models, the contributions of the three channels $\overline{K}^*(1410)^0\pi^+$, $\overline{K}^*_2(1430)^0\pi^+$ and $\overline{K}^*_3(1780)^0\pi^+$ are not significant, and their fit fractions are less than 0.2\%. Therefore, we remove them from the final model.  The final model (F) is composed of a non-resonant component and intermediate resonances modes, including $K^0_S\rho(770)^+$, $K^0_S\rho(1450)^+$, $\overline{K}^*(892)^0\pi^+$, $\overline{K}^*_0(1430)^0\pi^+$, $\overline{K}^*(1680)^0\pi^+$ and $\overline{\kappa}^0\pi^+$. The projections of the fit and the Dalitz plot can be found in Fig. \ref{fig:final}.

A deviation of efficiency between data and MC simulation will cause a deviation of the fit results. Therefore, a momentum-dependent correction is applied to the final results. The results are listed in the column \textquotedblleft Final\textquotedblright of Table \ref{tab:fitresult}.

In fits with these models, the formalism of the $\kappa$ is taken as the complex pole form, and the position of the pole $\kappa$ is allowed to float as a free complex parameter. The mass and width of the $K^*_0(1430)^0$, taken as a Breit-Wigner function, are also floated, since the measured values from E791 \cite{E791a} and CLEO-c \cite{CLEOc} in the $D^+\rightarrow K^-\pi^+\pi^+$ decay are not consistent with the PDG. Finally, it is measured that the pole of the $\kappa$ is at $(752\pm15\pm69^{+55}_{-73},-229\pm21\pm44^{+40}_{-55})$~MeV, which is consistent with the model C result of CLEO-c. And the mass and width of the $K^*_0(1430)^0$ are $1464\pm6\pm9^{+9}_{-28}$~MeV and $190\pm7\pm11^{+6}_{-26}$~MeV respectively, consistent with CLEO-c's results, while they are not consistent with the PDG. In the model without the $\overline{\kappa}$, the results are $1444\pm4$~MeV and $283\pm11$~MeV with statistical errors only, which are consistent with the PDG values.

\begin{figure}[htb]
  \centering
  \includegraphics[width=0.666\linewidth]{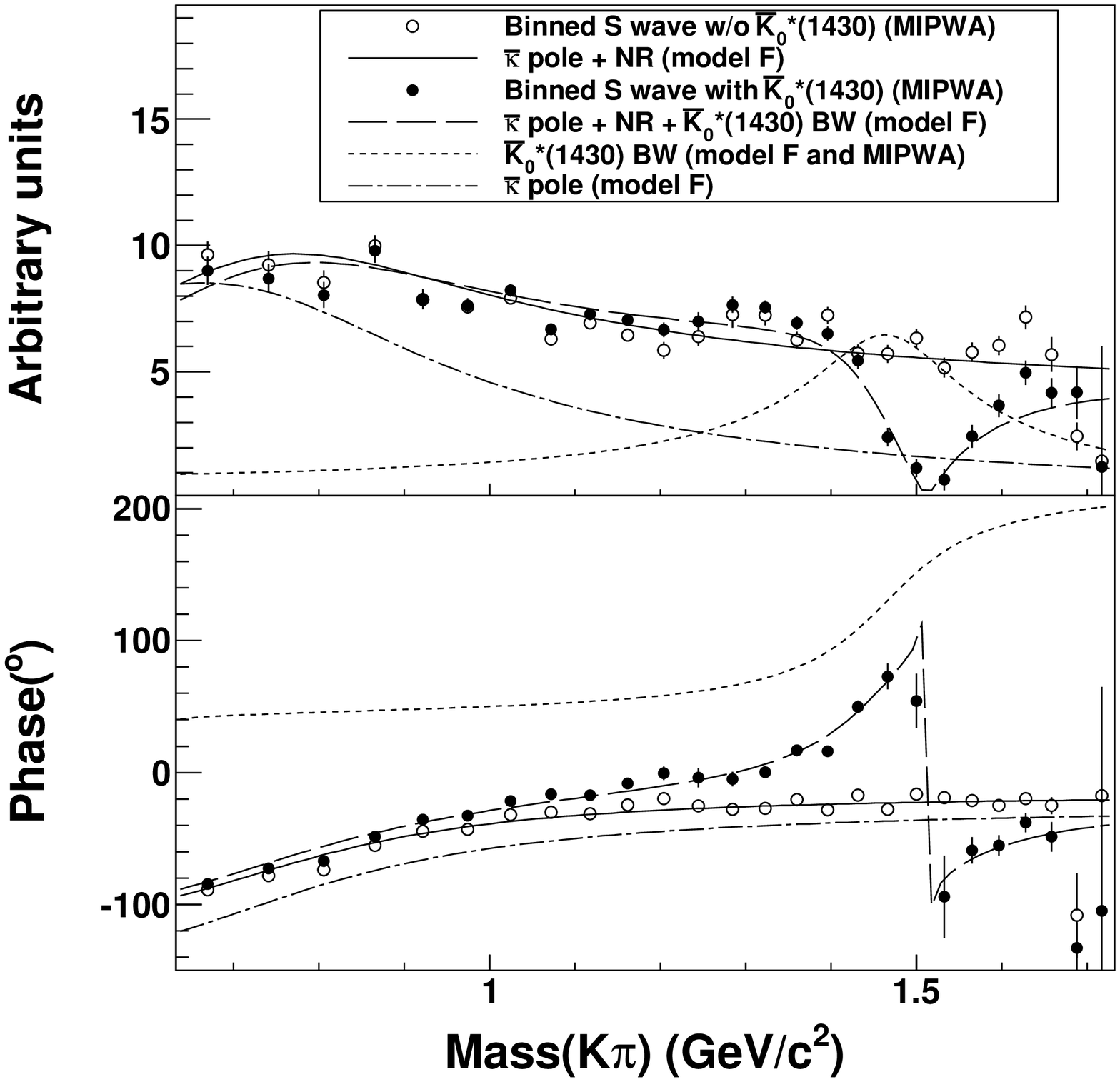}
  \begin{rotate}{45}
  \put(-100,250){\huge{BESIII Preliminary}}
  \end{rotate}
  \caption{The magnitude and phase of the $K\pi$ $S$ wave in model F and MIPWA. The blank dots with error bars for statistical uncertainties show the binned $K\pi$ $S$ wave without $\overline{K}^*_0(1430)$ and the black dots for the total $K\pi$ $S$ wave, respectively. Other curves show the $S$ wave components of model F.}
  \label{fig:mipwa}
\end{figure}

As cross-check, we perform a model-independent partial wave analysis (MIPWA) on the data, which is used in \cite{E791b}. The measured $S$-wave magnitude and phase of the $K^0_S\pi^0$ $S$-wave are demonstrated in Fig. \ref{fig:mipwa}. In order to compare with the previous $D^+\rightarrow K^-\pi^+\pi^+$ results, the magnitude and phase is changed to values relative to $\overline{K}^*(892)^0$. The results are consistent with the model-dependent analysis. It is obvious that there is still a phase shift in the $K^0_S\pi^0$ $S$-wave in the fit excluding the $\overline{K}^*_0(1430)^0$, which cannot be described with a non-resonant component, which indicates the $\overline{\kappa}$ is needed.

\section{Summary and Discussion}

\begin{table}[t]
  \centering
  \begin{tabular}{|l|c|}
  \hline
  \hline
  Mode & Partial Branching Fraction (\%) \\
  \hline
  \hline
  $D^+\rightarrow K^0_S\pi^+\pi^0$ Non Resonant & 0.32$\pm$0.05$\pm$0.25$_{-0.25}^{+0.21}$ \\
  $D^+\rightarrow \rho^+ K^0_S,\rho^+\rightarrow \pi^+\pi^0$  & 5.83$\pm$0.16$\pm$0.30$_{-0.15}^{+0.08}$ \\
  $D^+\rightarrow \rho(1450)^+ K^0_S,\rho(1450)^+ \rightarrow \pi^+\pi^0$  & 0.15$\pm$0.02$\pm$0.09$_{-0.11}^{+0.05}$ \\
  $D^+\rightarrow \overline{K}^*(892)^0\pi^+,\overline{K}^*(892)^0\rightarrow K^0_S\pi^0$ & $0.250\pm0.012\pm0.015_{-0.024}^{+0.022}$\\
  $D^+\rightarrow \overline{K}^*_0(1430)^0\pi^+,\overline{K}^*_0(1430)^0\rightarrow K^0_S\pi^0$& 0.26$\pm$0.04$\pm$0.05$_{-0.06}^{+0.03}$\\
  $D^+\rightarrow \overline{K}^*(1680)^0\pi^+,\overline{K}^*(1680)^0\rightarrow K^0_S\pi^0$&0.09$\pm$0.01$\pm$0.05$_{-0.08}^{+0.04}$\\
  $D^+\rightarrow \overline{\kappa}^0\pi^+,\overline{\kappa}^0\rightarrow K^0_S\pi^0$&0.54$\pm$0.09$\pm$0.28$_{-0.19}^{+0.14}$\\
  \hline
  NR+$\overline{\kappa}^0\pi^+$&1.30$\pm$0.12$\pm$0.12$_{-0.30}^{+0.11}$\\
  $K^0_S\pi^0$ $S$ wave & 1.21$\pm$0.10$\pm$0.16$_{-0.27}^{+0.05}$\\
  \hline
  \hline
  \end{tabular}
  \caption{The preliminary results of partial branching fractions calculated by combining our fit fractions with the PDG's $D^+\rightarrow K^0_S\pi^+\pi^0$ branching ratio. The errors shown are statistical, experimental systematic and modeling systematic respectively.}
  \label{tabsum}
\end{table}
BESIII has established the technology of Dalitz plot analysis. Based on it, the $D^+\rightarrow K^0_S \pi^+ \pi^0$ Dalitz plot is well-represented by a combination of a non-resonant component plus six quasi-two-body decays, $\overline{\kappa}$ included. The preliminary results are consistent with the results of E791 and CLEO-c in the $D^+\rightarrow K^-\pi^+\pi^+$ decay.

The final fit fraction and phase for each component, multiplied by the world average $D^+ \rightarrow K^0_S\pi^+\pi^0$ branching ratio of (6.99$\pm$0.27)\% \cite{PDG}, yield the partial branching fractions shown in Table \ref{tabsum}. The error on the world average branching ratio is incorporated by adding it in quadrature with the experimental systematic errors on the fit fractions to give the experimental systematic error on the partial branching fractions.

In this result, the $K^0_S\pi^0$ waves could be compared with the $K^-\pi^+$ waves in the $D^+\rightarrow K^-\pi^+\pi^+$ decay. For example, according to our measured branching ratio of $D^+\rightarrow \overline{K}^{*0}\pi^+\rightarrow K^0_S\pi^+\pi^0$ and the PDG value of branching ratio of $D^+\rightarrow \overline{K}^{*0}\pi^+\rightarrow K^-\pi^+\pi^+$ of (1.01$\pm$0.11)\%, the ratio of branching fraction of $D^+\rightarrow\overline{K}^{*0}\pi^+\rightarrow K^-\pi^+\pi^+$ and $D^+\rightarrow\overline{K}^{*0}\pi^+\rightarrow\overline{K}^0\pi^+\pi^0$ is calculated to $2.02\pm0.34$, which is consistent with what is expected.

\Acknowledgements
I am grateful to the National Natural Science Foundation of China under Contracts Nos. 11205178, in Institute of High Energy Physics of China, for supporting me the travel and the talk.

\end{document}